\documentstyle[seceq,epsf]{ptptex}
\def\lp{\stackrel{\leftarrow}{\partial}}
\def\rp{\stackrel{\rightarrow}{\partial}}
\def\lb{\{ \kern-.18em | }
\def\rb{| \kern-.18em \} }
\notypesetlogo  

\title{Phase-space Quantization of Field Theory }
\subtitle{{\footnotesize ANL-HEP-CP-99-06 \qquad Miami TH/1/99   }} 

\author{Cosmas {\sc Zachos }\footnote{Speaker. E-mail address: 
zachos@hep.anl.gov} 
and Thomas {\sc Curtright }$^{\S ,}$ \footnote{E-mail address: 
curtright@physics.miami.edu}   }
\inst{ High Energy Physics Division,
Argonne National Laboratory, Argonne, IL 60439-4815, USA \\  
$^{\S}$ Department of Physics, University of Miami,
Box 248046, Coral Gables, Florida 33124, USA } 
\recdate{March 1999}

\abst{ In this lecture, a limited introduction of gauge invariance in 
phase-space is provided, predicated on canonical transformations in 
quantum phase-space. Exact characteristic trajectories are also specified 
for the time-propagating Wigner phase-space distribution function: 
they are especially simple---indeed, classical---for
the quantized simple harmonic oscillator. This serves as the underpinning 
of the field theoretic Wigner functional formulation introduced. Scalar field 
theory is thus reformulated in terms of distributions in field
phase-space. This is a pedagogical selection from 
work published in \cite{base,cfz},
reported at the Yukawa Institute Workshop  
``Gauge Theory and Integrable Models", 26-29 January, 1999.} 

\begin{document}
\maketitle
\section{Introduction}

The third complete autonomous formulation of Quantum Mechanics, distinct from
conventional Hilbert space or path-integral quantization, is based on Wigner's
phase-space distribution function (WF), which is a special representation of
the density matrix \cite{wigner,moyal}. In this formulation, known as
deformation quantization \cite{bayen}, expectation values are computed by
integrating mere c-number functions in phase-space, the WF serving as a
distribution measure. Such phase-space functions multiply each other through
the pivotal $\star$-product \cite{groen}, which encodes the noncommutative
essence of quantization. The key principle underlying this quantization is the
$\star$-product's operational isomorphism \cite{bayen} to the conventional
Heisenberg operator algebra of quantum mechanics. 

Below, we address gauge invariance in phase-space through canonical
transformation to and from free systems. Further, we employ the $\star$-unitary
evolution operator, a ``$\star$-exponential", to specify the time propagation
of Wigner phase-space distribution functions. The answer is known to be
remarkably simple for the harmonic oscillator WF, and consists of classical
rigid rotation in phase-space for the full quantum system. It serves as the
underpinning of the generalization to field theory we consider, in which the
dynamics is specified through the evolution of c-number distributions on field
phase-space. We start by illustrating the basic concepts in 2d phase-space,
without loss of generality; then, in generalizing to field theory, we make the
transition to infinite-dimensional phase-space. 

Wigner functions are defined by 
\begin{equation}  
f(x,p)={\frac{1}{2\pi }}\int \!dy~\psi^{*}(x-{\frac{\hbar }{2}} 
y)~e^{-iyp}\psi(x+{\frac{\hbar }{2}}y)\;.
\end{equation} 
Even though they amount to spatial auto-correlation functions of 
Schr\"{o}dinger wavefunctions $\psi$, they can be determined without reference 
to such wavefunctions, in a logically autonomous structure. For instance, 
when the wavefunction is an energy ($E$) eigenfunction,
the corresponding WF is time-independent and satisfies the two-sided 
energy $\star $-genvalue equations\cite{dbfcam,cfz}, 
\begin{equation}  
H\star f=f\star H=E f \;,
\end{equation} 
where $\star $ is the essentially unique associative deformation of ordinary 
products on phase-space,
\begin{equation}
\star \equiv e^{{\frac{i\hbar }{2}}(\stackrel{\leftarrow }{\partial }_{x}%
\stackrel{\rightarrow }{\partial }_{p}-\stackrel{\leftarrow }{\partial }_{p}%
\stackrel{\rightarrow }{\partial }_{x})}\;.
\end{equation}
It is an exponentiation of the Poisson Bracket (PB) kernel, introduced by 
Groenewold \cite{groen} and developed in two important papers\cite{bayen}. 
Since it involves exponentials of derivative operators, it may be evaluated 
in practice through translation of function arguments, 
$f(x,p) \star g(x,p) = f(x+{i\hbar\over 2}\rp_p ,~ p-{i\hbar\over 2}\rp_x
)~g(x,p)$, to yield pseudodifferential equations. 

These WFs are real. They are bounded by the Schwarz inequality \cite{baker} to 
$-2/h\leq f\leq2/h$. They can go negative, and, indeed, they do for all but 
Gaussian configurations, negative values being a ready hallmark of interference.
Thus, strictly speaking, they are not probability distributions 
\cite{wigner}. However, upon integration over $x$ or $p$, they  yield
marginal probability densities in $p$ and $x$-space, respectively.
They can also be shown to be orthonormal\cite{dbfcam,cfz}. 
Unlike in Hilbert space quantum mechanics, naive superposition of solutions 
of the above does not hold, by virtue of Takabayashi's\cite{takabayashi,baker} 
fundamental nonlinear projection condition $f\star f=f/h$. 

Anyone beyond the proverbial jaded sophisticate should value the ready 
intuition on the shape of the WF based on the shape of the underlying
configuration $\psi(x)$.  Fix a point $x_0$, and reflect $\psi(x)$ around it,
$\psi(x_0 + (x-x_0)) \mapsto \psi(x_0-(x-x_0))$. Then, overlap this with the
starting configuration $\psi(x)$, to survey where the overlap vanishes, and
where it is substantial, thereby obtaining a qualitative picture of the support
of the WF at $x_0$. Thus, eg, it is evident by inspection that the WF $f(x,p)$
vanishes outside the absolute outer limits of support of the underlying
$\psi(x)$. However, a bimodal $\psi(x)$  (one consisting of two separated
bumps) will evidently yield $f(x,p)$ with nonvanishing support
(``interference") in the intermediate region between the two bumps in 
$\psi(x)$, even though $\psi(x)$ itself vanishes there.

\section{Gauge Systems}
While the question of adapting the WF formalism to gauge systems has been
addressed in the literature\cite{best}, and interesting proposals have been
made about accommodating nontrivial configurations involving Berry's
phase\cite{pati}, the problem has still not been settled in its full
generality. Here, the straightforward solution for gauge variation limited to a
merely 2d phase-space will be provided, by canonical mapping to a free
hamiltonian. This solution does not cover the more interesting topologically
nontrivial situations which arise in higher dimensions, and which still
languish in ambiguity. 
  
For notational simplicity, take $\hbar=1$ in this section.
The area element in phase-space is preserved by {\em canonical transformations} 
\begin{equation}
   (x,p) \mapsto (X(x,p), P(x,p)) ~,
\end{equation}
which yield trivial Jacobians ($dXdP= dx dp ~\{  X, P  \}$) by  
preserving the PBs, 
\begin{equation} 
\{u,v \}_{xp}\equiv {\partial   u \over \partial x  }
{\partial v    \over \partial p }-
{\partial u    \over \partial p  }
{\partial v   \over \partial x }~.
\end{equation}
They thus preserve the ``canonical invariants" of their functions,
$\{  X, P  \}_{xp} = 1$, hence  $\{  x, p  \}_{XP} = 1$. 
Equivalently, 
\begin{equation}
        \{  x, p  \}=\{  X, P  \},
\end{equation}
in any basis. Motion being a canonical transformation, Hamilton's 
classical equations of motion are preserved, 
for ${\cal H}(X,P)\equiv H(x,p)$, as well\cite{uematsu}. 
What happens upon quantization?

Since, in deformation quantization, the hamiltonian is a c-number 
function, and so transforms ``classically", ${\cal H}(X,P)\equiv H(x,p)$, 
the effects of a canonical transformation on the quantum theory 
will be carried by a ``covariantly" {\em  transformed Wigner function},
not the classical transform of it that is frequently demonstrated to be 
unworkable in the literature. Naturally, the answer can be deduced 
\cite{cfz} from Dirac's quantum transformation theory. 

Consider the classical canonical transformations specified by an arbitrary 
generating function $F(x,X)$:
\begin{equation}
 p={\partial F(x,X) \over \partial x } ~, \qquad \qquad 
 P=-{\partial F(x,X) \over \partial X }~.
\end{equation}
Following Dirac's celebrated exponentiation \cite{pamd} of such a 
generator, Scr\"{o}dinger's  energy eigenfunctions transform canonically 
through a generalization of the ``representation-changing" Fourier 
transform. Namely, 
\begin{equation}
 \psi _{E}(x)=N_{E}\int dX\,e^{iF(x,X)}\,\Psi _{E}(X) ~.
\end{equation}
Thus, the pair of Wigner functions in the respective canonical variables, 
$f(x,p)$ and 
\begin{equation} 
 {\cal F}(X,P)={1\over 2\pi} \int\! dY ~\Psi^* (X-{Y\over2} )~e^{-iYP} 
 \Psi(X+{Y\over 2} ),    \label{tfmwf} 
\end{equation}
are connected through a convolution by a transformation functional, 
\begin{equation}    
 f(x,p)=\int \!dX\int \!dP ~{\cal T} (x,p;X,P)~{\cal F}(X,P)\;.  \label{convol} 
\end{equation}
(In proof, it has come to our attention that this equation has already been 
given in Ref \cite{garcia}). In Curtright et al.\cite{cfz}, this 
transformation functional was evaluated to be 
\begin{equation}
 {\cal T}(x,p;X,P)
 =\frac{|N|^2}{2\pi} \!\int\!dY dy~ \exp \left(\! -iyp+iYP\!-\!
 iF^{*}(x-{\frac{y}{2}},X-{\frac{Y}{2}}) 
 +iF(x+{\frac{y}{2}} ,X+{\frac{Y}{2}})\right ).\label{babba}
\end{equation}

Now consider the succession of two canonical transformations with 
generating functionals $F_1$ and $F_2$, respectively\cite{uematsu},
useful for generating all other classes of canonical transformations,
\begin{equation}
 (x,p) \mapsto (X, P) \mapsto ({\chi }, {\varpi} ). 
\end{equation}

By inspection of (\ref{convol},\ref{babba}), it follows that the corresponding
transformation functional ${\cal T}(x,p;{\chi},{\pi})$ is 
\begin{equation}
{\cal T}(x,p;{\chi,\varpi})=
 \!\int\!dXdP~{\cal T}_1(x,p;X,P) ~{\cal T}_2(X,P;\chi,\varpi)
  \qquad  \qquad  \qquad  \qquad  \qquad  
\end{equation}
$$
 =\frac{|N_1 N_2|^2}{(2\pi)^2} \!\int\!dX dP dY dydw dW~
  \exp ( -iyp+iYP -iwP +iW{\varpi}\qquad  \qquad  
$$
$$
\qquad   -iF^{*}_1(x-{\frac{y}{2}},X-{\frac{Y}{2}}) 
+iF_1(x+{\frac{y}{2}},X+{\frac{Y}{2}})
-iF^{*}_2(X-{\frac{w}{2}},{\chi}-{\frac{W}{2}}) 
+iF_2(X+{\frac{w}{2}} ,{\chi}+{\frac{W}{2}}) ),
$$
and, performing the trivial $P$, whence $w$ integrations,
$$
=\frac{|N|^2}{2\pi} \!\int\!dy dW~ \exp \left ( -iyp+iW{\varpi} 
 -i{\sf F}^{*}(x-{\frac{y}{2}},{\chi-}{\frac{W}{2}}) 
 +i{\sf F}(x+{\frac{y}{2}} ,{\chi}+{\frac{W}{2}})
 \right ),
$$
where the effective generator {\sf F} is specified by Dirac's integral 
expression:
\begin{equation}
\exp (i {\sf F}(x,{\chi}))=
\!\int\!dX  ~ \exp \left (  iF_1(x,X)+iF_2(X,{\chi}) \right ).\label{compo}
\end{equation}
The classical result for the concatenation of two canonical transformations
would require that the exponent of the integrand does not depend on $X$,
so that the momenta $P$ at the intermediate phase space point coincide. 
Instead, quantum mechanically, Dirac's celebrated result dictates 
integration over all intermediate points, as seen. 

A succession of several intermediate points $(X_i,P_i)$ integrated over
naturally generalizes to entire paths of canonical transformations; their 
effective generators entering in (\ref{babba}) are given by the usual
Dirac path integral expressions generalizing the above. For motion, all
generators are the same function, \linebreak 
$F=S(x=x(t_i), X=x(t_f))$,
the action increment. Thus, the generalization yields the familiar  
Dirac/Feynman functional integral propagator. Consequently,
each wavefunction in (\ref{tfmwf}) propagates independently by 
its respective Feynman effective action. 

Finally, consider the canonical transformation from the free hamiltonian 
$H=p^2/2m$ to the one for a minimally coupled particle in a ``magnetic" field, 
${\cal H}=(\varpi -eA(\chi ))^2/2m$, in one space dimension.
In the formalism outlined, this is effected by a succession of two canonical 
transformations, 
\begin{equation}
(x,p) \mapsto (X=-p, P=x) \mapsto \left ( {\chi }=x, {\varpi}=p+eA(x)\right ), 
\end{equation}
generated by $F_1=-xX$ and $F_2=X{\chi}- e \int^{\chi}dz ~  A(z) $,
respectively. 

Consequently, the effective exponentiated generator (\ref{compo}) is 
\begin{equation}
\exp (i {\sf F}(x,{\chi}))=2 \pi 
\delta(x-{\chi}) ~  \exp \left (-  ie \int^x dz A(z) \right ). 
\end{equation}
Thus, the gauge-invariant eigenfunction $\psi$ of the free hamiltonian
transforms to the {\em gauge-variant} eigenfunction $\Psi$ of the 
EM hamiltonian,
\begin{equation}
\psi(x) = e^{-ie \int^x dz A(z)}   ~ \Psi(x) ~.
\end{equation}
Hence, the transformation functional (\ref{babba}) reduces to  
\begin{equation}
{\cal T}(x,p;X,P)=\frac{1}{2\pi} 
\int\!dY \exp\! \left ( -iY(p-P) -ie 
\int^{x+Y/2}_{x-Y/2} \!\! dz A(z) \right ) ~ \delta (x-X). 
\end{equation}

As a result, the free-particle WF canonically transforms to the 
{\em gauge-invariant} one for a particle in the presence of a vector potential,
already considered by \cite{best} (with $\hbar$ reinstated for completeness): 
\begin{equation}
f(x,p)={1\over 2\pi}\int\! dy~\Psi^* (x-{\hbar\over2} y )~e^{-iyp 
-(ie/\hbar)  \int^{x+\hbar y/2}_{x-\hbar y/2} dz A(z)} ~
\Psi(x+{\hbar\over2} y).    
\end{equation}
Of course, this gauge-invariant WF is controlled by the Moyal equation 
(detailed in the next section) driven by the free $H$, not ${\cal H}$,
which controls ${\cal F}$ instead. 

The above discussion is only a starting point of 
setting up gauge invariance in phase space. 
By contrast to this discussion, in more than one space dimensions, 
the line integral of the vector potential is ambiguous, and straight paths 
from every point {\boldmath $x$} to all endpoints
(involving {\boldmath $y$}, integrated over) are inequivalent, in general, to 
alternate paths in nontrivial field settings. One should then not expect 
these problems to be canonically equivalent to free ones. 
The motivated reader is referred to Pati\cite{pati}.

\section {Time Evolution}

Time-dependence for WFs was succinctly specified as flows in phase-space by 
Moyal through the commutator bracket\cite{moyal} bearing his name, 
\begin{equation} 
i\hbar \frac{\partial }{\partial t}\,f(x,p;t)=H\star f(x,p;t)-f(x,p;t)\star H.
\label{jomo}
\end{equation}
This turns out to be the essentially unique associative 
generalization\cite{vey} of the PB, to which it reduces as 
$\hbar\rightarrow 0$, yielding Liouville's  theorem of classical mechanics, 
${\partial}_t \,f + \{ f,H \} =0$ \footnote{ 
The question was asked at the talk what generalizes the $\star$- product when 
redundant variables are present, notably for Dirac Brackets (DB); but it has 
not been answered. Eg, on a hypersphere $S^n$, if the redundant 
variables are eliminated, one deals with simple coordinate changes for 
PB, as usual. But if, instead,  the redundant 
variables are retained, subject to the constraint 
$x_0^2+ x_1^2+ x_2^2+...+ x_n^2=1$,
and hence $x_0 p_0+ x_1 p_1+ x_2 p_2+...+ x_n p_n =0$,
the PBs are supplanted by the DBs suitable to this isospin 
hypersphere\cite{corrigan}, 
$$
\lb x_i, x_j \rb =0, ~~~~\lb x_i, p_j \rb =\delta_{ij} - x_i x_j, 
~~~ \lb p_i, p_j \rb = x_j p_i- x_i p_j~. 
$$
The kernel realizing these brackets is 
$$
\bowtie ~ =  \lp_x\cdot\rp_p - \lp_x\cdot x ~ x\cdot\rp_p -\lp_p\cdot\rp_x 
    + \lp_p\cdot x ~  x\cdot\rp_x+\lp_p\cdot p~ x\cdot\rp_p-\lp_p\cdot x~ 
       p\cdot \rp_p.
$$
All exponentiation assignments of this kernel examined so far have not 
yielded an associative modification of the $\star$-product.
For a different approach, see Antonsen\cite{antonsen}.} .  

For the evolution of the fundamental phase-space 
variables $x$ and $p$, time evolution is simply the convective part 
of Moyal's equation, so the apparent sign is reversed, while the Moyal Bracket 
actually reduces to the PB. 
That is, the $\hbar$-dependence
drops out, and these variables, in fact, evolve simply by the 
{\em classical} Hamilton's equations of motion, $\dot{x}= \partial_p H$,
$\dot{p}= -\partial_x H$. 

What is the time-evolution of a WF like? This is the first question 
answered through the isomorphism\cite{bayen} of $\star$-multiplication 
associative combinatorics to the parallel algebraic 
manipulations of quantum mechanical operators, which are familiar.
Eqn (\ref{jomo}) can be solved for the time-trajectories of 
the WF, which turn out to be notably simple.
By virtue of the $\star$-unitary evolution operator, 
a ``$\star$-exponential"\cite{bayen},
\begin{equation} 
U_{\star} (x,p;t)=e_\star ^{itH/\hbar} \equiv 
1+(it/\hbar)H(x,p) + {(it/\hbar)^2\over 2!}  H\star H +{(it/\hbar )^3\over 3!} 
 H\star H\star H +...,
\end{equation}
the time-evolved Wigner function is obtainable formally in 
terms of the Wigner function at $t=0$ through associative combinatoric
operations completely analogous to the conventional formulation of quantum 
mechanics of operators in Hilbert space.
Specifically, 
\begin{equation} 
f(x,p;t)=U_{\star}^{-1} (x,p;t) \star f(x,p;0)\star U_{\star}(x,p;t) .    
\label{evol}
\end{equation}
As indicated, the dynamical variables evolve classically, 
\begin{equation}  
{dx \over dt}= {x \star H-H\star x \over  i\hbar}= \partial_p H ~,
\end{equation} 
and 
\begin{equation}  
{dp \over dt}= {p \star H-H\star p \over i\hbar}=- \partial_x H ~.
\end{equation} 
Consequently, by associativity, the quantum evolution,
\begin{equation}  
x(t)=U_{\star} \star x \star U_{\star}^{-1}, 
\end{equation} 
\begin{equation}  
p(t)=U_{\star} \star p \star U_{\star}^{-1}, 
\end{equation} 
turns out to flow along {\em classical} trajectories. 

What can be said about this formal time-evolution expression?
If the WF can be written as a $\star$-function, ie a sum of $\star$-products
of the phase-space variables, then associativity will permit application of
the above $\star$-similarity transformation throughout the WFs. 

Any WF in phase-space, upon Fourier transformation resolves to 
\begin{equation} 
f(x,p)= \int\! da db ~\tilde{f}(a,b) ~ e^{iax} e^{ibp}.
 \end{equation} 
However, note that exponentials of individual functions of 
$x$ and $p$ are also $\star$-exponentials of the same functions, or 
 $\star$-versions of these functions, since the $\star$-product trivializes
in the absence of a conjugate variable, so that 
\begin{equation}
e^{iax} ~e^{ibp}=e_{\star} ^{iax} ~e_{\star} ^{ibp}  . \label{grolp}
\end{equation} 
Moreover, this is proportional to a $\star$-product, since  
\begin{equation}  
e_{\star} ^{iax}\star  e_{\star} ^{ibp}=
e_{\star} ^{ia(x+i\hbar \rp_p/2)} ~ e_{\star} ^{ibp}
=e_{\star} ^{iax} e_{\star} ^{ibp} e^{-i\hbar ab / 2}. \label{grulp}
\end{equation} 

Consequently, any Wigner function can be written as 
\begin{equation} 
f(x,p)= \int\! da db ~\tilde{f}(a,b) ~e^{i\hbar ab/ 2} ~
e_{\star} ^{iax}\star  e_{\star} ^{ibp}.
\end{equation} 
It follows then, that, by insertion of $U_{\star}\star U_{\star}^{-1}  $ 
pairs at every $\star$-multiplication, in general, 
$$
f(x,p;t)=\int\! da db ~\tilde{f}(a,b) ~e^{{i\hbar ab /2}} ~
e_{\star} ^{iaU_{\star}^{-1}   \star x \star U_{\star}}\star   ~  
e_{\star} ^{ibU_{\star}^{-1}  \star p \star U_{\star}} 
$$
\begin{equation} 
=\int\! da db ~\tilde{f}(a,b) ~e^{i\hbar ab /2}~
e_{\star} ^{iax(-t)}\star  e_{\star} ^{ibp(-t)}.
\end{equation} 

Unfortunately, in general, the above steps cannot be simply reversed to yield 
an integrand of the  form $\tilde{f}(a,b) ~ e^{iax(-t)} e^{ibp(-t)}$. 
But, in some limited fortuitous circumstances, they can, and in this 
case the evolution
of the Wigner function reduces to merely backward evolution of its arguments 
$x,p$ along classical trajectories, while its functional form itself remains 
unchanged:
\begin{equation} 
f(x,p;t)=f\left( x(-t),p(-t);0\right).  \label{gold}
 \end{equation} 

To illustrate this, consider the simple linear harmonic oscillator 
(taking $m=1$, $\omega=1$),
\begin{equation}  
H={p^2 +x^2\over 2}= \frac{x-ip} {\sqrt{2}} \star \frac{x+ip} {\sqrt{2}} +
{\hbar\over 2} ~.\label{simplex}
\end{equation} 
It is easily seen that 
\begin{equation}  
i\hbar \dot{x}= x \star H-H\star x = i\hbar p ~ , \qquad 
i\hbar \dot{p}= p \star H-H\star p = -i\hbar x   ~,
\end{equation} 
and thus the canonical variables indeed evolve classically: 
\begin{eqnarray}  
X\equiv x(t)&=&U_{\star}\star x\star U_{\star}^{-1}= x\cos t + p \sin t, 
\nonumber \\  
P\equiv p(t)&=&U_{\star}\star p\star U_{\star}^{-1}= p\cos t - x \sin t.  
\label{rotation} 
\end{eqnarray} 
This also checks against the explicitly evaluated 
$\star$-exponential for the    SHO\cite{bayen}, 
$e_\star ^{itH/\hbar}=\frac{1} {\cos (t/2)} \exp ( {2i\tan (t/2)} H/{\hbar} )$.

Now, recall the degenerate reduction of the Baker-Campbell-Hausdorff 
combinatoric identity for any two operators with {\em constant} commutator 
with respect to any associative multiplication, thus for
any phase-space functions $\xi$ and $\eta$ under $\star$-multiplication. If
\begin{equation} 
\xi \star \eta - \eta \star \xi=c,
 \end{equation} 
then, 
\begin{equation} 
e_{\star} ^{\xi}\star  e_{\star} ^{\eta}=
e_{\star} ^{\xi+\eta} ~e^{c/2}.
\end{equation} 
Application of this identity as well as (\ref{grulp}) and (\ref{grolp}) 
yields directly
\begin{eqnarray} 
e_{\star}^{iax(-t)} \star e_{\star} ^{ibp(-t)} e^{i\hbar ab/2}&  =&
e_{\star}^{i(a\cos t+b\sin t ) x+ i(b\cos t-a\sin t)p }\nonumber \\
&=& e_{\star}^{i(a\cos t+b\sin t ) x } \star e_{\star} ^{i(b\cos t-a\sin t)p } 
e^{i\hbar (a\cos t+b\sin t) (b\cos t-a \sin t)/2}  \nonumber \\
&=& e_{\star}^{i(a\cos t+b\sin t ) x } e_{\star} ^{i(b\cos t-a\sin t)p}
\nonumber \\
&=& e^{i(a\cos t+b\sin t ) x }~e^{i(b\cos t-a\sin t)p }~.  
 \end{eqnarray} 
Consequently,
\begin{equation} 
f(x,p;t)= \int\! da db ~\tilde{f}(a,b) ~ e^{iax(-t)} e^{ibp(-t)},
 \end{equation} 
and hence the reverse convective flow (\ref{gold}) indeed obtains for the SHO.

The result for the SHO is the preservation of the functional form of the Wigner 
distribution function along classical phase-space trajectories:
\begin{equation}  
f(x,p;t)=f(x \cos t - p \sin t,     p \cos t + x \sin t  ;0). \label{preserv}
\end{equation} 
What this means is that {\em any} Wigner distribution rotates 
uniformly on the phase plane around the origin, essentially classically,
even though it provides a complete quantum mechanical description.

\vskip 0.2cm

\epsf{file=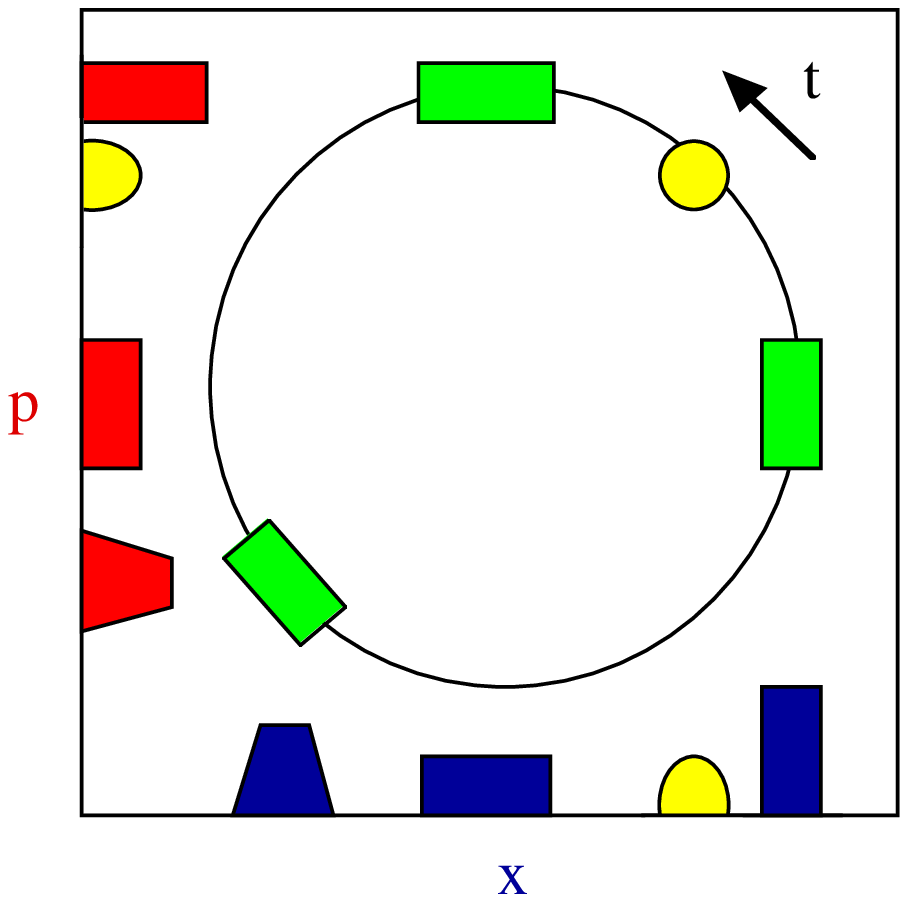,width=9cm}   

Naturally, this rigid rotation in phase-space preserves areas, and
thus illustrates the uncertainty principle. By contrast, in general, in the 
conventional formulation of quantum mechanics, this result is deprived of 
intuitive import, or, at the very least, simplicity: upon integration in $x$ 
(or $p$) to yield usual probability densities, the rotation induces 
apparent complicated shape variations of the oscillating probability density 
profile, such as wavepacket spreading (as evident in the shadow 
projections on the $x$ and $p$ axes of the figure). Only when 
(as is the case for coherent states \cite{almeida}) a Wigner function 
configuration has an 
{\em additional} axial $x-p$ symmetry around its {\em own} center, 
will it possess an invariant profile upon this rotation, and hence a 
shape-invariant oscillating probability density. (In the figure, a rectangle is
taken to represent a generic configuration, and a small circle to represent 
such a coherent state.)

The result (\ref{preserv}), of course, is not new. It was clearly recognized 
by Wigner \cite{kim}. 
It follows directly from (\ref{jomo}) for (\ref{simplex}) that 
\begin{equation}  
\left(  \partial_t +p \partial_x- x \partial_p\right) f(x,p;t)=0~. 
\label{character} 
\end{equation} 
The characteristics of this partial differential equation correspond to the 
above uniform rotation in phase space, so it is easily seen to be solved 
by (\ref{preserv}).   The result was given 
explicitly in \cite{groen} and also \cite{bartlett}, following
different derivations. Lesche \cite{lesche},
has also reached this result in an elegant fifth derivation, 
by noting that for quadratic Hamiltonians such as this one,
the linear rotation of the dynamical variables (\ref{rotation}) 
leaves the symplectic quadratic form invariant, and thus the $\star$-product 
invariant. That is, the gradients in the $\star$-product may also be taken to 
be with respect to the time-evolved canonical variables (\ref{rotation}),
$X$ and $P$; 
hence, after inserting $U_{\star}\star U_{\star}^{-1}  $ in the 
$\star$-functional form of $f$, the $\star$-products may be taken 
to be with respect to $X$ and $P$, and the functional form of $f$ is preserved, 
(\ref{gold}). This only holds for quadratic Hamiltonians, which thus generate
{\em linear} canonical transformations. 

Dirac's interaction representation may then be based on this property,
for a general Hamiltonian consisting of a basic SHO part, 
$H_0=(p^2 +x^2)/2$, supplemented by an interaction part,
\begin{equation}
H= H_0 + H_I.
\end{equation} 
Now, upon defining
\begin{equation}
w\equiv e_\star ^{itH_0/\hbar} \star f \star e_\star ^{-itH_0/\hbar},
\end{equation} 
it follows that Moyal's evolution equation reads,
\begin{equation}
i\hbar \frac{\partial }{\partial t}\,w(x,p;t)
={\cal H}_I \star w (x,p;t)- w(x,p;t)\star {\cal H}_I,
  \end{equation} 
where 
${\cal H}_I \equiv e_\star ^{itH_0/\hbar}\star H_I \star 
e_\star ^{-itH_0/\hbar}$. Expressing $H_I$ as a $\star$-function leads to 
a simplification. 

In terms of the convective variables (\ref{rotation}), $X,P$,
${\cal H}_I(x,p) =H_I (X,P)$, and $w(x,p;t) =f(X,P;t)$, while 
$\star$ may refer to these convective variables as well. Finally, then,
\begin{equation}
i\hbar \frac{\partial }{\partial t}\,f(X,P;t)
= H_I(X, P) \star f (X, P;t)- f(X, P;t)\star H_I (X, P).  \label{interaction}
\end{equation} 
In the uniformly rotating frame of the convective variables, the WF 
time-evolves according to the interaction Hamiltonian---while, for vanishing 
interaction Hamiltonian,  $f (X, P;t)$ is constant and yields 
(\ref{preserv}). Below, in generalizing to field theory, 
this provides the basis of the interaction picture of perturbation theory,
where the canonical fields evolve classically as above \footnote{
One may note alternate discussions\cite{nachbag,antonsen} of field theoretic 
interaction representations in phase-space, which do not appear coincident 
with the one to be presented here.}.

\section{Scalar Field Theory in Phase Space}
 
To produce Wigner functionals in scalar field theory, one may start from the 
standard, noncovariant, formulation of field theory in Hilbert space, 
in terms of Schr\"{o}dinger wave-functionals. 

For a free field Hamiltonian, the energy eigen-functionals are Gaussian in 
form. For instance, without loss of generality, in two dimensions ($x$ is a 
spatial coordinate, and $t=0$ in all fields),
the ground state functional is 
\begin{equation}
\Psi[ \phi] =\exp\left(  -\frac{1}{2\hbar}\int dx\,\phi\left(
x\right)  \sqrt{m^{2}-\nabla_{x}^{2}}~ \phi\left(  x\right)  \right).
\end{equation}
Boundary conditions are assumed such that the $\sqrt{m^{2}-\nabla_{x}^{2}}$
kernel in the exponent is naively self-adjoint. 
``Integrating by parts'' one of the $\sqrt{m^{2}-\nabla_{z}^{2}}$ kernels,
functional derivation $\delta\phi\left(  x\right)  /\delta\phi\left(  z\right)
=\delta\left(  z-x\right) $ then leads to  
\begin{equation}  
\hbar\frac{\delta}{\delta\phi\left(  z\right)  }\Psi [\phi] 
=-\left(  \sqrt{m^{2}-\nabla_{z}^{2}}\,\phi\left(  z\right)  \right)
\Psi[ \phi] ,
\end{equation} 

\begin{equation}  
\hbar^{2}\frac{\delta^{2}}{\delta\phi\left( w \right) \delta\phi
\left( z\right)  }\Psi[ \phi] =
\end{equation} 
$$
=\left(  \sqrt{m^{2}-\nabla_{w}^{2}}\,
\phi\left(  w\right)  \right)  \left(  \sqrt{m^{2}-\nabla_{z}^{2}}\,
\phi\left(  z\right)  \right)  \,\Psi[ \phi] -\hbar\sqrt
{m^{2}-\nabla_{z}^{2}}\,\delta\left(  w-z\right)  \,\Psi[ \phi]. 
$$

Note that the divergent zero-point energy density, 
\begin{equation}  
E_0= {\hbar\over 2  }    \lim_{w\rightarrow z} 
\sqrt {m^{2}-\nabla_{z}^{2}}~ \delta\left(  w-z\right), 
\end{equation} 
may be handled rigorously using $\zeta$-function regularization. 

Leaving this zero-point energy present leads to the 
standard energy eigenvalue equation, again through integration by parts, 
\begin{equation}  
\frac{1}{2} \int dz\,\left(  -\hbar^{2}\frac{\delta^{2}}
{\delta\phi\left(  z\right)  ^{2} }+\phi\left(  z\right)  
\left(  m^{2}-\nabla_{z}^{2}\right)  \phi\left( z\right)  \right)  \Psi [\phi ] 
=E_0 ~\Psi[ \phi]. 
\end{equation} 

A natural adaptation to the corresponding Wigner functional is the following.
For a functional measure $\left[ d\eta / 2\pi \right]  =\prod_{x} 
d\eta\left(  x\right)  / 2\pi$, one obtains
\begin{equation}
W [\phi,\pi] =\int\!\left[  \frac{d\eta}{2\pi}\right]
~\Psi^{\ast}\left[  \phi-{\frac{\hbar}{2}\eta}\right]  \,
e^{-i\int dx\,\eta\left(  x\right)  \pi\left(  x\right)  }~
\Psi\left[  \phi+{\frac{\hbar }{2}\eta}\right] ,
\end{equation}
where $\pi\left(  x\right)  $ is to be understood as the local field variable
canonically conjugate to $\phi\left(  x\right)  $. However, in this
expression, both $\phi$ and $\pi$ are {\em classical} variables, not
operator-valued fields, in full analogy to the phase-space quantum mechanics 
already discussed. 

For the Gaussian ground-state wavefunctional, this evaluates to 
\begin{eqnarray}
W\left[  \phi,\pi\right]   & =&\!\int\left[  \frac{d\eta}{2\pi}\right]
\exp\left(  -\frac{1}{2\hbar}\int dx\,\left(  \phi\left(  x\right)
-{\frac{\hbar}{2}\eta}\left( x\right) \right) \sqrt{m^{2}-\nabla_{x}^{2} 
}\,\left(  \phi\left(  x\right)  -{\frac{\hbar}{2}\eta}\left(  x\right)
\right)  \right) \times \nonumber \\
&\times &e^{-i\int dx\,\eta\left(  x\right)  \pi\left(  x\right)  } 
\exp\left(  -\frac{1}{2\hbar}\int dx\,\left(  \phi\left(  x\right)
+{\frac{\hbar}{2}\eta}\left(  x\right)  \right) \sqrt{m^{2}-\nabla_{x}^{2}}
\left( \phi\left( x\right) +{\frac{\hbar}{2}\eta}\left( x\right)\right) 
\right) \nonumber \\
&=&\exp\left( -\frac{1}{\hbar}\int dx\,\phi\left(  x\right)  \,\sqrt
{m^{2}-\nabla_{x}^{2}}\,\phi\left( x\right) \right) \times \nonumber \\
&\times& \left(  \int\left[  \frac{d\eta}{2\pi}\right]  \,e^{-i\int
dx\,\eta\left(  x\right)  \pi\left(  x\right)  }\exp\left(  -\frac{\hbar}{4}
\int dx\,{\eta}\left(  x\right)  \,\sqrt{m^{2}-\nabla_{x}^{2}}\,{\eta
}\left(  x\right)  \right)  \right).
\end{eqnarray}
So
\begin{equation}
W[\phi,\pi] ={\cal N}\exp\left(  -\frac{1}{\hbar}\int 
dx\,\left(  \left(  \phi\left(  x\right)  \,\sqrt{m^{2}-\nabla_{x}^{2}}\,
\phi\left(  x\right)\right) + \left(  {\pi}\left(  x\right)  \,\left(
\sqrt{m^{2}-\nabla_{x}^{2}}\right) ^{-1}\! {\pi}\left(  x\right)  \right) 
\right) \right) ,
\end{equation}
where ${\cal N}$ is a normalization factor. It is the expected collection of 
harmonic oscillators. 

This Wigner functional is, of course \cite{cfz}, an energy
$\star$-genfunctional, also checked directly. For
\begin{equation}  
H_0[\phi,\pi]\equiv \frac{1}{2} \int dx\left(  \pi\left(  x\right) ^{2}
+\phi\left( x\right)  \left( m^{2}-\nabla_{x}^{2}\right)  
\phi\left(  x\right)  \right),
\end{equation} 
and the predictable infinite dimensional phase-space generalization\footnote
{Recall that $x$ is now a labelling parameter, not a phase-space variable.} 
\begin{equation}
\star \equiv \exp \left( {\frac{i\hbar }{2}}\int dx~ \left( 
\stackrel{\leftarrow } {\delta\over\delta\phi (x)} 
\stackrel{\rightarrow }{\delta\over\delta\pi (x)}
-\stackrel{\leftarrow }{\delta\over\delta\pi (x)}
\stackrel{\rightarrow }{\delta\over\delta\phi (x)}\right)  \right) ,
\end{equation}  
it follows that 
\begin{eqnarray}
&&H_0\star W=  \nonumber \\
  &=&\!\int \frac{dx}{2}\!\left(
\left(  \pi\left(  x\right)  -\frac{i\hbar}{2}\frac{\delta}{\delta\phi\left(
x\right)  }\right)^{2} \!+\left(  \phi\left(  x\right)  
+\frac{i\hbar}{2}\frac{\delta}{\delta
\pi\left(  x\right)  }\right)  \left(  m^{2}-\nabla_{x}^{2}\right)  \left(
\phi\left(  x\right)  +\frac{i\hbar}{2} \frac{\delta}{\delta\pi\left(
x\right)  }\right)\right)  W\left[  \phi,\pi\right]  \nonumber \\  
& =&\int\frac{dx }{2}  \left(
\pi\left(  x\right)  ^{2}-\frac{\hbar^{2}}{4}\frac{\delta}{\delta\pi\left(
x\right)  }\left(  m^{2}-\nabla_{x}^{2}\right)  \frac{\delta}{\delta\pi\left(
x\right)  } 
+\phi\left(  x\right)  \left(  m^{2}-\nabla_{x}^{2}\right)  \phi\left(
x\right)  -\frac{\hbar^{2}}{4}\frac{\delta^{2}}{\delta\phi\left(  x\right)
^{2}} \right)  W\left[  \phi,\pi\right]  \nonumber\\
&=& E_0 ~W\left[\phi ,\pi\right].
\end{eqnarray}
This is indeed the ground-state Wigner energy-$\star$-genfunctional. 
The $\star$-genvalue is again the zero-point energy, which could have been 
removed by point-splitting the energy density, as indicated earlier. 
There does not seem to be a simple point-splitting procedure that 
regularizes the star product as defined above and also preserves associativity.

As in the case of the SHO discussed above, free-field time-evolution for 
Wigner functionals is also effected by Dirac
delta functionals whose support lies on the classical field time evolution
equations. 
Fields evolve according to the equations,
\begin{equation}  
-i\hbar\partial_{t}\phi  =H\star\phi-\phi\star H,\qquad \qquad 
-i\hbar\partial_{t}\pi  =H\star\pi-\pi\star H.
\end{equation} 
For $H_0$, these equations are the classical evolution
equations for free fields, 
\begin{equation}  
\partial_{t}\phi\left(  x,t\right)     =\pi\left(  x,t\right) , \qquad \qquad 
\partial_{t}\pi\left(  x,t\right)  =-\left(  m^{2}-\nabla_{x}^{2}\right)
\phi\left(  x,t\right).
\end{equation} 

Formally, the solutions are represented as 
\begin{equation}
\label{prp1}  
\phi\left(  x,t\right)  =\cos\left(  t\sqrt{m^{2}-\nabla_{x}^{2}}\right)
\phi\left(  x,0\right)  +\sin\left(  t\sqrt{m^{2}-\nabla_{x}^{2}}\right)
\frac{1}{\sqrt{m^{2}-\nabla_{x}^{2}}}~ \pi\left(  x,0\right)   
\end{equation}
\begin{equation}
\label{eulalie}  
\pi\left(  x,t\right)     =-\sin\left(  t\sqrt{m^{2}-\nabla_{x}^{2}}\right)
\sqrt{m^{2}-\nabla_{x}^{2}}~\phi\left(  x,0\right)  +\cos\left(  t\sqrt
{m^{2}-\nabla_{x}^{2}}\right)  \pi\left(  x,0\right).  
\end{equation}
From these, it follows by the functional chain rule that
$$
\int dx\left(  \pi\left(  x,0\right)  \frac{\delta}{\delta\phi\left(
x,0\right)  }-\left(  \left(  m^{2}-\nabla_{x}^{2}\right)  \phi\left(
x,0\right)  \right)  \,\frac{\delta}{\delta\pi\left(  x,0\right)  }\right)
\qquad \qquad \qquad \qquad 
$$
\begin{equation}  
=\int dx\left(  \pi\left(  x,t\right)  \frac{\delta}{\delta\phi\left(
x,t\right)  }-\left(  \left(  m^{2}-\nabla_{x}^{2}\right)  \phi\left(
x,t\right)  \right)  \,\frac{\delta}{\delta\pi\left(  x,t\right)  }\right)
\end{equation} 
for any time $t$. 

Consider the free-field Moyal evolution equation for a generic (not necessarily 
energy-$\star$-genfunctional) WF, corresponding to (\ref{character}), 
\begin{equation}
\partial_{t}W  = -\int dx\,\left(  \pi\left(  x\right)  \frac{\delta}{\delta
\phi\left(  x\right)  }-\phi\left(  x\right)  \left(  m^{2}-\nabla_{x}%
^{2}\right)  \frac{\delta}{\delta\pi\left(  x\right)  }\right)  W.
\end{equation}
The solution is an infinite-dimensional version of (\ref{gold}),
\begin{equation}  
W\left[  \phi,\pi;t\right]  =W\left[  \phi\left(  -t\right)  ,\pi\left(
-t\right)  ;0\right].
\end{equation} 

Adapting the method of characteristics for
first-order equations to a functional context,
one may simply check this solution again using the chain rule for functional 
derivatives, and the
field equations {\em evolved backwards in time} as specified:
$$
\partial_{t}W\left[  \phi,\pi;t\right]     =\partial_{t}W\left[  \phi\left(
-t\right)  ,\pi\left(  -t\right)  ;0\right]  
\qquad\qquad\qquad\qquad\qquad\qquad\qquad\qquad\qquad\qquad
$$ 
\begin{eqnarray}
& =&\int dx\,\left(  \partial
_{t}\phi\left(  x,-t\right)  \,\frac{\delta}{\delta\phi\left(  x,-t\right)
}+\partial_{t}\pi\left(  x,-t\right)  \,\frac{\delta}{\delta\pi\left(
x,-t\right)  }\right)  
W\left[  \phi\left(  -t\right)  ,\pi\left(  -t\right);0\right]  \nonumber \\
& =&\int dx\left(  \,\left(  -\pi\left(  x,-t\right)  \right)  \frac{\delta
}{\delta\phi\left(  x,-t\right)  }+\left(  \left(  m^{2}-\nabla_{x}%
^{2}\right)  \phi\left(  x,-t\right)  \right)  \,\frac{\delta}{\delta
\pi\left(  x,-t\right)  }\right)  W\left[  \phi\left(  -t\right)  ,\pi\left(
-t\right)  ;0\right]\nonumber   \\
& =&\int dx\left(  \,\left(  -\pi\left(  x,-t\right)  \right)  \frac{\delta
}{\delta\phi\left(  x,-t\right)  }+\left(  \left(  m^{2}-\nabla_{x}%
^{2}\right)  \phi\left(  x,-t\right)  \right)  \,\frac{\delta}{\delta
\pi\left(  x,-t\right)  }\right)  W\left[  \phi,\pi;t\right] \nonumber \\
& =&-\int dx\,\left(  \pi\left(  x\right)  \frac{\delta}{\delta\phi\left(
x\right)  }-\left(  m^{2}-\nabla_{x}^{2}\right)  \phi\left(  x\right)
\,\frac{\delta}{\delta\pi\left(  x\right)  }\right)  W\left[  \phi
,\pi;t\right ].
\end{eqnarray}
The quantum Wigner Functional for free fields time-evolves along classical  
field configurations (which propagate noncovariantly according to 
(\ref{prp1},\ref{eulalie})).
In complete analogy to the interaction 
representation for single particle quantum mechanics, (\ref{interaction}), the 
perturbative series in the interaction Hamiltonian (written as a 
$\star$-function of fields) is then defined
in terms of the convective free field variables $\Phi,\Pi$ propagating 
classically.
Specifically, 
\begin{equation}
i\hbar \frac{\partial }{\partial t}\,W[\Phi,\Pi;t]
= H_I[\Phi,\Pi] \star W [\Phi,\Pi ;t]- W[\Phi,\Pi ;t]\star 
H_I [\Phi,\Pi]~.  
\end{equation} 

\section*{Acknowledgements}
This work was supported in part by NSF grant PHY 9507829, and
by the US Department of Energy, Division of High Energy Physics, 
Contract W-31-109-ENG-38. The speaker, CZ, thanks R Sasaki and T Uematsu for 
their kind hospitality and support at the Yukawa Institute 
Workshop; its participants for their helpful questions; and A Sasaki for
assistance with graphics.

\end{document}